\documentclass[preprint]{aastex}

\usepackage{epsfig}
\usepackage{natbib}     
\usepackage{verbatim}   
\usepackage{graphicx}   
\usepackage{amsmath,amsthm,amsfonts,amsopn,amssymb} 

\usepackage{longtable,pdflscape}   
\usepackage{lscape}
\usepackage{longtable}
\usepackage{afterpage}
\usepackage{rotating}	
\usepackage{ulem}

\shorttitle{Single Transits and Eclipses Observed by K2}
\shortauthors{LaCourse \& Jacobs}

\begin{document}


\title{Single Transits and Eclipses Observed by K2}


\author{
Daryll M. LaCourse\altaffilmark{1},
Thomas L. Jacobs\altaffilmark{2}
}
\slugcomment{Accepted to RNAAS Feb. 6, 2018}

\altaffiltext{1}{7507 52nd Place NE Marysville, WA, 98270; daryll.lacourse@gmail.com, USA}
\altaffiltext{2}{12812 SE 69th Place Bellevue, WA 98006; tomjacobs128@gmail.com,
USA}



\keywords{Planets and satellites: detection -- surveys -- catalogs -- binaries: eclipsing}

Photometric survey data from the {\it Kepler} (Koch, et al. 2010) mission have been used to discover and characterize thousands of transiting exoplanet and eclipsing binary (EB) systems. These discoveries have enabled empirical studies of occurrence rates\footnote{For a review of occurence rate studies see: \url{https://exoplanetarchive.ipac.caltech.edu/docs/occurrence\_rate\_papers.html}} which reveal that exoplanets are ubiquitous and found in a wide variety of system architectures and physical compositions. Because the detection strategy of {\it Kepler} is most sensitive to short orbital periods, the vast majority of these objects reside within 1 AU of their host star. Although other detection techniques have successfully identified exoplanets at wider orbits beyond the snow lines of their respective host stars (e.g., radial velocity, microlensing, direct imaging), occurrence rates within this population remain poorly constrained. As such, identifying long period objects (LPOs) from archival {\it Kepler} and {\it K2} (Howell, et al. 2014) data is valuable from both a statistical and theoretical standpoint, particularly for massive gas giants which are thought to heavily influence the formation and evolution dynamics of their respective systems. Here we present a catalog of 164 single transit and eclipse candidates detected during a comprehensive survey of all currently available {\it K2} data.

Identification of LPO candidates from Kepler data has been accomplished in several previous studies, which we briefly summarize here. Wang, et al. (2013) and Wang et al. (2015) leveraged a crowd sourcing strategy based on manual inspections performed by citizen-scientists of the Planet Hunters project (Fischer, et al. 2012), identifying 15 double-transit candidates and 17 single-transit candidates. Similarly, Uehara, et al. (2016) performed a visual inspection of 7557 Kepler Objects of Interest which recovered 28 single transits; including 7 objects consistent with Neptune-sized and Jupiter-sized exoplanets. Kipping, et al. (2014) and Kipping et al. (2015) utilized an automated pipeline to identify and validate a Uranus-sized exoplanet orbiting Kepler-421 and a Jupiter-sized exoplanet orbiting Kepler-167. Finally, Foreman-Mackey, et al. (2016) utilized a pipeline equipped with a probabilistic model comparison function to systematically search the light curves of 39,036 G and K dwarfs observed by Kepler, recovering 3 double-transit candidates and 13 single-transit candidates. An important result from the Foreman-Mackey study was to show an estimated occurrence rate for G/K dwarfs of 2.00(+/-0.7) exoplanets smaller than Jupiter in the 2 to 25 year orbital period range. 

The search for LPOs has continued successfully into the {\it K2} mission. Indeed, the first confirmed exoplanet of the mission, K2-2b, was identified from a single transit recovered during the mission's engineering test phase (Vanderburg, et al. 2015). Subsequently, an automated search was performed by Osborn, et al. (2016) for selected stars from Campaigns 1,2 and 3, which recovered 7 additional LPOs. 

Motivated by these results, we undertook a visual survey of 288,399 unique stellar light curves derived from {\it K2} Campaigns 0 through 14. For Campaigns 0, 1 and 2 we downloaded Target Pixel Files from the \textsc{MAST}\footnote{https://archive.stsci.edu/kepler} and extracted light curves in an automated fashion with Guest Observer software \textsc{PyKE}\footnote{http://pyke.keplerscience.org}(Still \& Barclay 2012; Vinícius, et al. 2017). For Campaigns 3 through 14, both the \textsc{K2 Ames} and \textsc{K2SFF} light curves were made available at the MAST and both sets of data were surveyed independenty by each author. For all data sets, candidates for which the LPO signal contaminated multiple EPIC targets or were noted to be associated with a spacecraft resaturation event, were discarded as false positives. Surviving targets are organized in Table 1 by Campaign and in order of decreasing brightness. Measurements for midpoint, depth, duration were taken with the light curve examination software \textsc{LcTools}\footnote{https://sites.google.com/a/lctools.net/lctools/home}. Included stellar parameters are derived from the EPIC catalog (Huber, et al. 2016), except for Campaign 0 where we include Teff estimations from the \textsc{K2-TESS} catalog\footnote{https://filtergraph.com/tess\_k2campaigns}(Stassun, et al. 2014).

Based purely on an assessment of transit depth, it is tempting to conclude that many of the signals are caused by long period EBs. However we have opted to present the sample as an ensemble, without conclusively discriminating between stellar and sub-stellar companions. Although the radii of the LPOs can be inferred from a single transit, such estimations are only as accurate as our knowledge of the host star's parameters; for the vast majority of our sample such characterization has not been performed via asteroseismology or spectral fitting. An additional consideration is that large numbers of both late-type and evolved stars have been observed in K2, which can allow scenarios where an exoplanet transit mimics an EB, or vice versa. As such, we do not provide putative radius and period estimations here, save to note that if the EPIC parameters are assumed to be wholly accurate, 38 of the LPOs seem consistent with a planetary origin. Additional photometric or RV follow up is encouraged to properly elucidate the nature of these candidates.

{\it Note Added in Review}: After submission of this research letter, we became aware of a large number of additional unpublished single transits in Hugh Osborn’s PhD thesis\footnote{https://warwick.ac.uk/fac/sci/physics/research/astro/publications/phd\_msc/hughosborn.phd.pdf}. We are gratified to see a significant overlap between the two samples.

\acknowledgments

We are grateful to Andrew Vanderburg, Saul Rappaport, Joeseph Schmitt, Fei Dai and Ben Montet for helpful discussions and comments during the preparation of this catalog. We thank LcTools author Allan R. Schmitt for facilitating this research by making his software available for our survey. This letter includes data from the K2 mission, which is funded by the NASA Science Mission directorate. Much of the data presented was obtained from the Mikulski Archive for Space Telescopes (MAST), which is governed by STScI and operated by the Association of Universities for Research in Astronomy Inc., under NASA contract NAS526555. We thank all past and present members of the {\it Kepler} \& {\it K2} teams for their efforts to ensure the fantastic success of both missions.



Facilities: \facility{Kepler/K2}



\begin{center}
    
\textbf{References}

\end{center}

Auvergne, M., Bodin, P., Boisnard, L., et al. 2009, A\&A, 506, 411

Fischer, D. A., Schwamb, M. E., Schawinski, K., et al. 2012, MNRAS, 419, 2900 

Foreman-Mackey, D., Morton, T. D., Hogg, D. W., Agol, E. \& Scholkopf, B., 2016, arXiv:1607.08237

Huber D., et al., 2016, ApJS, 224, 2

Koch D. G., et al., 2010, ApJ, 713, L79

Kipping, D. M., Torres, G., Buchhave, L. A., et al. 2014b, ApJ, 795, 25 

Kipping, D. M., Torres, G., Henze, C., et al. 2016, ApJ, 820, 112

Osborn, H. P., Armstrong, D. J., Brown, D. J. A., et al. 2016, MNRAS, 457, 2273

Stassun, K. G., Pepper, J. A., Paegert, M., De Lee, N., \& Sanchis-Ojeda, R. 2014, arXiv:1410.6379

Still, M., \& Barclay, T. 2012, Astrophysics Source Code Library, ascl:1208.004

Vinícius, Z., Barentsen, G., Hedges, et al. 2018, Zenodo, doi:10.5281/zenodo.835583

Uehara, S., Kawahara, H., Masuda, K., Yamada, S.,  Aizawa, M. 2016, ApJ, 822, 2

Vanderburg A., Johnson J. A., 2014, PASP, 126, 948

Wang, J., Fischer, D. A., Barclay, T., et al. 2013, ApJ, 776, 10

Wang, J., Fischer, D. A., Barclay, T., et al. 2015, ApJ, 815, 127


















\begin{landscape}

\begin{deluxetable}{lcccccccc}
\tabletypesize{\scriptsize}

\tablecaption{Single Transits and Eclipses Observed by K2\label{tab:singles}}
\tablewidth{0pt}
\tablehead{
\colhead{EPIC} & 
\colhead{Campaign}   & 
\colhead{$K_{\rm P}$} & 
\colhead{$T_{\rm eff}$} & 
\colhead{R*} & 
\colhead{BJD$_0$}    &
\colhead{Depth}   &
\colhead{Duration} &
\colhead{Comments}  \\
\colhead{} & 
\colhead{} &
\colhead{(mag)} & 
\colhead{(K)} & 
\colhead{R$_{\odot}$}    &
\colhead{(BJD-2454833)}    &
\colhead{(ppm)} &
\colhead{(hours)}    &
\colhead{}  
}
\startdata

248811085	& 14	&	 11.25	&	6327	&	1.53	&	3115.3795	&	91337	&	6.3746	&		\\
248555345	& 14	&	 11.60	&	5192	&	6.07	&	3106.6452 	&	125978	&	15.6912	&	Deep	\\
248854690	& 14	&	 11.98	&	5462	&	2.58	&	3117.3517	&	84973	&	25.4987	&		\\
248847494	& 14	&	 12.17	&	4931	&	0.79	&	3134.1158 	&	2236	&	56.3914	&		\\
201854636	& 14	&	 12.19	&	4936	&	3.98	&	3118.2002 	&	30324	&	16.1819	&		\\
248749087	& 14	&	 13.80	&	4812	&	0.65	&	3094.5298 	&	1833	&	5.8824	&		\\
248657359	& 14	&	 13.92	&	4790	&	0.70	&	3093.0281 	&	19297	&	14.2224	&		\\
248607265	& 14	&	 15.14	&	4383	&	0.56	&	3091.0972 	&	19639	&	5.8844	&		\\
201792207	& 14	&	 15.35	&	4101	&	0.48	&	3105.8900 	&	465848	&	3.9229	&	Deep; secondary eclipse	\\
248912804	& 14	&	 16.59	&	3334	&	0.17	&	3126.7501 	&	94658	&	9.8064	&		\\
248873506	& 14	&	 17.30	&	3917	&	0.35	&	3108.1574 	&	114537	&	11.7696	&		\\
247692298	& 13	&	 9.71	&	10398	&	3.20	&	3006.2056 	&	188904	&	10.2960	&	Deep	\\
246732631	& 13	&	 10.77	&	8224	&	2.06	&	3003.3757 	&	137259	&	21.5760	&	Deep; secondary eclipse		\\
247756662	& 13	&	 11.80	&	4202	&	35.34	&	3049.91875 	&	28528	&	310.8888	&	Flat-bottom	\\
247762843	& 13	&	 11.96	&	6527	&	1.39	&	3002.60955 	&	1300	&	5.3952	&		\\
246849982	& 13	&	 12.52	&	9715	&	1.85	&	3061.15675	&	235239	&	26.4792	&	Deep	\\
246696483	& 13	&	 12.72	&	6149	&	1.72	&	3022.59145 	&	32984	&	17.1624	&		\\
247865067	& 13	&	 12.94	&	5013	&	8.33	&	3042.9614 	&	2266	&	8.3352	&		\\
247795097	& 13	&	 12.99	&	3606	&	166.45	&	3001.9245 	&	192036	&	12.7488	&	Deep	\\
247594337	& 13	&	 13.63	&	4908	&	6.90	&	2997.6037	&	175064	&	12.2592	&	Deep	\\
247967714	& 13	&	 13.75	&	5076	&	2.01	&	3012.0791 	&	258337	&	4.9032	&	Deep	\\
247450113	& 13	&	 14.04	&	5886	&	1.86	&	3003.5390 	&	3363	&	14.7120	&		\\
246906371	& 13	&	 14.55	&	4912	&	5.74	&	2994.9265 	&	24957	&	7.3536	&	Deep	\\
247814074	& 13	&	 14.64	&	6245	&	1.76	&	3050.1637	&	443535	&	11.7696 &	Deep; secondary eclipse		\\
247015294	& 13	&	 15.86	&	6267	&	1.26	&	2997.8078 	&	174621	&	9.3168	&		\\
246194159	& 12	&	 9.48	&	6340	&	1.81	&	2914.3878 	&	378934	&	8.82665	&	Deep; secondary eclipse	\\
246359551	& 12	&	 11.14	&	4696	&	5.33	&	2914.3570 	&	204068	&	21.0858	&	Deep; secondary eclipse		\\
245929407	& 12	&	 11.64	&	-	    &	-	    &	2927.1473 	&	22554	&	6.374717	&		\\
245964933	& 12	&	 12.18	&	5480	&	0.88	&	2925.2779 	&	40201	&	10.7879	&		\\
246394998	& 12	&	 12.52	&	5877	&	1.12	&	2913.9893 	&	438858	&	5.3940 &	Deep; secondary eclipse	\\
246089124	& 12	&	 13.05	&	5865	&	1.10	&	2941.4699	&	138529	&	9.3167	&	Deep; secondary eclipse		\\
246361128	& 12	&	 14.38	&	4717	&	0.68	&	2911.2000 	&	45500	&	6.8651	&		\\
246163364	& 12	&	 14.89	&	4681	&	0.73	&	2972.8118 	&	69921	&	34.8155	&	Deep; secondary eclipse; flat-bottom	\\
246301164	& 12	&	 14.97	&	5063	&	0.78	&	2925.7785 	&	131436	&	4.4132	&	Deep	\\
246425172	& 12	&	 15.16	&	4866	&	4.64	&	2939.1922 	&	62953	&	67.6696	&	Deep; secondary eclipse; flat-bottom	\\
246302531	& 12	&	 15.56	&	4146	&	0.49	&	2914.9906 	&	63915	&	4.4133	&		\\
246020606	& 12	&	 15.69	&	5274	&	0.81	&	2939.7230 	&	34977	&	5.8842	&		\\
246331715	& 12	&	 16.95	&	5429	&	0.81	&	2927.6786 	&	57229	&	6.3747 &		\\
246279882	& 12	&	 17.52	&	5158	&	6.79	&	2939.5191	&	65549	&	53.9395	&	Deep; secondary eclipse; flat-bottom	\\
230887315	& 11	&	 8.47	&	5262	&	2.74	&	2827.3093 	&	30968	&	8.3352	&		\\
224937601	& 11	&	 10.47	&	9696	&	5.30	&	2869.8795 	&	202073	&	44.1312	&	Deep	\\
223332021	& 11	&	 10.74	&	-   	&	-   	&	2887.6860	&	72754	&	17.1624	&	Deep; secondary eclipse		\\
226600397	& 11	&	 10.85	&	6744	&	3.58	&	2862.2176 	&	53438	&	43.6440	&	Deep; secondary eclipse		\\
232334247	& 11	&	 11.01	&	6566	&	1.59	&	2831.1303 	&	1998	&	18.1435 &		\\
231674968	& 11	&	 11.28	&	6073	&	3.37	&	2839.8854	&	10914	&	46.0941	&		\\
236157481	& 11	&	 11.87	&	7525	&	3.09	&	2880.6772 	&	116832	&	22.0662	&	Deep	\\
231312005	& 11	&	 11.94	&	5854	&	1.60	&	2852.5122 	&	4908	&	8.8272	&		\\
235099239	& 11	&	 13.21	&	4046	&	0.47	&	2856.5988 	&	2200	&	14.7096	&		\\
240410914	& 11	&	 13.44	&	4799	&	40.26	&	2837.1685 	&	589868	&	10.7880	&	Deep	\\
240323152	& 11	&	 14.61	&	8545	&	3.39	&	2841.3162	&	12643	&	5.8848 &		\\
242209485	& 11	&	 14.67	&	6229	&	2.98	&	2861.1554 	&	39727	&	11.7672	&		\\
224703312	& 11	&	 15.15	&	5509	&	2.98	&	2882.4858 	&	146172	&	11.7672	&	Deep; secondary eclipse	\\
229021605	& 10	&	 10.47	&	4895	&	6.00	&	2785.7140 	&	17416	&	14.2204	&		\\
201132684	& 10	&	 11.67	&	5549	&	0.87	&	2797.8904 	&	2196	&	12.7512	&	Apparent multi-planet system	\\
201092629	& 10	&	 11.85	&	5259	&	0.76	&	2778.0310 	&	1126	&	5.8843	&		\\
201479221	& 10	&	 11.91	&	5697	&	0.98	&	2787.3679 	&	159819	&	10.7879	&	Deep	\\
228786343	& 10	&	 12.72	&	6140	&	1.19	&	2814.7579 	&	14966	&	4.4132	&		\\
228804202	& 10	&	 13.45	&	5719	&	0.94	&	2790.4029 	&	731	&	3.9240	&		\\
229022237	& 10	&	 13.60	&	5168	&	1.84	&	2751.6236 	&	21467	&	9.3170	&		\\
201093731	& 10	&	 13.74	&	6134	&	1.27	&	2801.8443 	&	35729	&	6.3747	&	Deep	\\
228891397	& 10	&	 13.76	&	5088	&	6.36	&	2780.2586 	&	246193	&	44.6227	&	Deep	\\
201510813	& 10	&	 14.19	&	5739	&	0.93	&	2802.0174 	&	149321	&	5.8843	&	Deep	\\
201496916	& 10	&	 17.72	&	3833	&	0.33	&	2799.5966 	&	221704	&	10.2975	&	Deep	\\
220315458	& 8	&	 11.59	&	5832	&	1.05	&	2611.4965 	&	985	&	4.9056	&		\\
220186865	& 8	&	 12.00	&	5158	&	0.87	&	2571.6956 	&	1973	&	5.8848	&		\\
220562610	& 8	&	 12.51	&	5915	&	1.40	&	2563.6864 	&	2813	&	11.7672	&		\\
220606084	& 8	&	 13.00	&	5761	&	1.50	&	2618.2492 	&	4928	&	8.3352	&		\\
220152847	& 8	&	 13.19	&	5129	&	0.80	&	2568.8657 	&	9191	&	12.2592	&		\\
220605820	& 8	&	 13.76	&	5089	&	3.47	&	2590.8710 	&	13884	&	6.3744	&		\\
220515668	& 8	&	 14.38	&	5385	&	0.82	&	2599.4419 	&	152383	&	4.9032	&	Deep	\\
220208795	& 8	&	 14.25	&	5064	&	0.75	&	2575.9658 	&	225648	&	6.8664	&	Deep	\\
215067200	& 7	&	 5.92	&	9210	&	4.32	&	2493.8114 	&	151215	&	41.6806	&	Flat-bottom	\\
216831785	& 7	&	 11.08	&	6251	&	1.34	&	2505.0904 	&	406	&	23.0472	&		\\
213832800	& 7	&	 11.24	&	5326	&	2.77	&	2488.4585 	&	81263	&	8.3361	&	Deep	\\
218751675	& 7	&	 11.44	&	-	    &	-	    &	2502.2907 	&	3592	&	11.2782	&	Deep; secondary eclipse	\\
215307988	& 7	&	 11.55	&	7426	&	-   	&	2489.3981 	&	410977	&	17.1626	&	Deep; secondary eclipse	\\
213332545	& 7	&	 11.79	&	5892	&	1.16	&	2505.5805	&	108103	&	4.4132	&		\\
218187050	& 7	&	 12.29	&	5379	&	8.74	&	2512.0469 	&	2216	&	14.7107	&		\\
214783208	& 7	&	 12.86	&	4206	&	38.41	&	2493.7603 	&	7104	&	25.9896	&		\\
215894766	& 7	&	 13.06	&	5267	&	2.36	&	2501.4119 	&	26419	&	6.3746	&		\\
213867148	& 7	&	 13.14	&	5150	&	16.29	&	2533.8786 	&	8750	&	98.5620	&		\\
219240689	& 7	&	 14.12	&	4953	&	8.97	&	2532.6731	&	25884	&	42.6611	&	Flat-bottom	\\
212549089	& 6	&	 11.50	&	5986	&	1.33	&	2412.0360 	&	53688	&	14.2224	&		\\
212820423	& 6	&	 11.95	&	4820	&	5.10	&	2433.3465 	&	6805	&	22.0656	&		\\
212542155	& 6	&	 12.52	&	5800	&	0.98	&	2389.6836	&	1132	&	8.3352	&		\\
212685467	& 6	&	 13.26	&	5714	&	0.91	&	2409.6665 	&	9319	&	7.3560	&		\\
212694013	& 6	&	 13.36	&	4913	&	5.03	&	2409.1754 	&	7113	&	40.6992	&		\\
212555615	& 6	&	 13.42	&	5606	&	2.11	&	2428.6783 	&	2358	&	3.9240	&		\\
212805678	& 6	&	 13.69	&	4986	&	4.27	&	2460.5206 	&	60463	&	23.0472	&	Secondary eclipse	\\
212325089	& 6	&	 13.69	&	4983    &	4.85	&	2441.2127 	&	115890	&	53.9376	&	Flat-bottom	\\
212343520	& 6	&	 13.80	&	5100	&	0.88	&	2451.0103	&	80532	&	6.8664	&	Secondary eclipse	\\
212477236	& 6	&	 13.96	&	5023	&	3.25	&	2449.5901	&	4389	&	15.2016	&		\\
212554009	& 6	&	 14.56	&	5834	&	0.96	&	2462.2474 	&	141582	&	25.4976	&	Deep	\\
212332380	& 6	&	 14.56	&	5213	&	0.86	&	2449.8354 	&	227278	&	14.2201	&	Deep; secondary eclipse	\\
212775521	& 6	&	 14.65	&	5252	&	2.28	&	2424.7963 	&	12666	&	6.8640	&		\\
212732378	& 6	&	 14.66	&	5272	&	0.65	&	2396.7732	&	58214	&	6.3768	&		\\
212715204	& 6	&	 16.67	&	4193	&	0.47	&	2399.3379 	&	185073	&	4.9032	&		\\
212152316	& 5	&	 10.31	&	5631	&	2.29	&	2374.2193 	&	66344	&	30.4032	&		\\
211633458	& 5	&	 10.72	&	4830	&	10.55	&	2368.6513 	&	1197	&	9.3168	&		\\
211598816	& 5	&	 10.75	&	7437	&	2.19	&	2363.3709	&	9718	&	3.9240	&		\\
211924561	& 5	&	 10.75	&	6976	&	1.17	&	2346.8923	&	2144	&	3.4320 &		\\
212026444	& 5	&	 10.81	&	5938	&	1.18	&	2343.3680	&	644	&	25.4976	&	Flat-bottom	\\
211953574	& 5	&	 11.31	&	6027	&	1.54	&	2360.5608 	&	590	&	6.3744	&		\\
211351543	& 5	&	 11.38	&	6251	&	1.66	&	2373.7598 	&	491	&	6.3744	&		\\
211939692	& 5	&	 11.75	&	6404	&	1.44	&	2337.5035 	&	1204	&	11.7696	&	Apparent multi-planet system	\\
211939692	& 5	&	 11.75	&	6404	&	1.44	&	2359.9068 	&	710	&	9.3168	&	Apparent multi-planet system	\\
211892898	& 5	&	 11.77	&	4787	&	8.74	&	2356.3519	&	8984	&	62.2756	&		\\
211498244	& 5	&	 11.88	&	3731	&	0.31	&	2321.5460	&	94154	&	6.8664	&		\\
212070574	& 5	&	 12.10	&	6039	&	0.91	&	2376.5789 	&	3843	&	5.3928	&		\\
211351097	& 5	&	 12.32	&	6214	&	1.42	&	2328.8308 	&	905	&	7.3560	&		\\
212011230	& 5	&	 12.37	&	4856	&	4.15	&	2332.4882 	&	5003	&	6.3744	&		\\
211821192	& 5	&	 12.58	&	5769	&	0.95	&	2344.1743 	&	1103	&	7.3553	&		\\
211840710	& 5	&	 12.72	&	5162	&	6.88	&	2353.1026 	&	3890	&	7.3560	&		\\
211894612	& 5	&	 12.76	&	6063	&	1.31	&	2349.9258 	&	44324	&	8.8272	&		\\
211490542	& 5	&	 12.90	&	6134	&	1.37	&	2343.1733 	&	80800   &   10.2960	&	Deep; secondary eclipse	\\
211995462	& 5	&	 12.96	&	5771	&	0.97	&	2370.2761 	&	256342	&	11.7696	&	Deep	\\
212012030	& 5	&	 13.07	&	5127	&	0.81	&	2341.4668 	&	202281	&	4.9032	&	Deep	\\
211503363	& 5	&	 13.22	&	4988	&	4.34	&	2364.5142 	&	4616	&	23.5368	&		\\
211390677	& 5	&	 13.31	&	4886	&	4.86	&	2318.7983	&	6780	&	39.7195	&	Higher SnR in Ames aperture	\\
211411112	& 5	&	 13.40	&	6026	&	1.25	&	2345.9526 	&	26062	&	4.4112	&		\\
211489484	& 5	&	 15.99	&	3767	&	0.342	&	2359.6206 	&	109099	&	4.4112	&	Deep; secondary eclipse	\\
210725198	& 4	&	 10.39	&	7899	&	2.31	&	2258.1411	&	103537	&	18.6336	&	Deep; secondary eclipse	\\
211087003	& 4	&	 11.64	&	6072	&	1.25	&	2251.0917	&	8089	&	7.8456	&	Apparent multi-planet system	\\
210857749	& 4	&	 12.70	&	6392	&	1.46	&	2251.9701 	&	184683	&	13.7304	&	Deep	\\
211064647	& 4	&	 14.91	&	5613	&	0.74	&	2273.9654 	&	13344	&	6.3744	&		\\
210823406	& 4	&	 15.74	&	4102	&	0.49	&	2257.3639	&	174979	&	5.8848	&	Deep	\\
210760314	& 4	&	 16.20	&	4470	&	0.58	&	2248.1500 	&	274496	&	8.8272	&		\\
210825751	& 4	&	 16.43	&	3865	&	0.39	&	2275.9263 	&	10907	&	15.2016	&		\\
210843533	& 4	&	 16.94	&	3566	&	0.24	&	2256.1794 	&	64606	&	9.8064	&		\\
211075893	& 4	&	 17.72	&	4431	&	0.57	&	2253.7886 	&	163591	&	11.7696	&		\\
205966706	& 3	&	 9.86	&	6200	&	2.73	&	2184.7830 	&	119208	&	7.8457	&	Deep	\\
206253908	& 3	&	 11.18	&	6864	&	1.75	&	2150.6618	&	100778	&	6.8651	&       \\
205936222	& 3	&	 12.09	&	5581	&	2.00	&	2152.7766 	&	325665	&	6.3748	&	Deep	\\
206008070	& 3	&	 13.82	&	4983	&	4.88	&	2172.0842	&	8626	&	97.5815	&	Flat-bottom	\\
203914123	& 2	&	 9.03	&	4801	&	11.11	&	2113.2841 	&	26917	&	75.5344	&	Noted in Osborn, et al. (2016)	\\
204546592	& 2	&	 9.93	&	4848	&	10.92	&	2066.965 	&	10212	&	20.5968	&		\\
203865172	& 2	&	 10.67	&	6199	&	1.90	&	2082.5250 	&	6043	&	2.4504	&		\\
203746451	& 2	&	 11.50	&	5627	&	0.93	&	2114.0612 	&	5550	&	11.7672	&		\\
204918110	& 2	&	 11.80	&	6442	&	2.03	&	2111.8748 	&	112761	&	22.5552	&	Deep; flat-bottom	\\
203311200	& 2	&	 11.89	&	6787	&	1.94	&	2121.0070 	&	3656	&	13.7304	&	Noted in Osborn, et al. (2016)	\\
204533587	& 2	&	 12.24	&	5717	&	0.91	&	2093.4343 	&	1350	&	9.3168 &		\\
204634789	& 2	&	 12.29	&	6029	&	2.02	&	2088.0105	&	7814	&	2.9424	&	Noted in Osborn, et al. (2016)	\\
204086428	& 2	&	 12.31	&	6149	&	1.43	&	2066.6689	&	2160	&	6.3744	&		\\
203011840	& 2	&	 12.48	&	5063	&	1.91	&	2080.2971 	&	475216	&	19.1256	&	Deep	\\
204952800	& 2	&	 12.79	&	6328	&	1.22	&	2110.6789 	&	428538	&	6.3744	&	Deep	\\
205272592	& 2	&	 14.51	&	4544	&	12.57	&	2102.3539	&	338711	&	12.7488	&	Deep	\\
204100531	& 2	&	 14.58	&	6384	&	1.80	&	2098.7673 	&	1436692	&	16.1808	&	Deep	\\
204776782	& 2	&	 14.77	&	7247	&	1.95	&	2127.0038 	&	267132	&	14.2201 &	Deep	\\
201207683	& 1	&	 10.64	&	5293	&	9.15	&	2002.3225 	&	236704	&	7.3554	&	Deep; Noted in Schmitt, et al. (2016)	\\
201775904	& 1	&	 11.57	&	6103	&	1.58	&	2002.5464 	&	14162	&	5.3939	&		\\
201631267	& 1	&	 12.78	&	5012	&	2.64	&	1996.6831 	&	6207	&	7.3554	&	Noted in Osborn, et al. (2016)	\\
201176672	& 1	&	 13.98	&	4542	&	0.50	&	2044.7690	&	31203	&	16.6723	&		\\
201720401	& 1	&	 14.74	&	5024	&	3.79	&	1983.3515 	&	23941	&	36.2876	&	Noted in Osborn, et al. (2016)	\\
201663371	& 1	&	 15.05	&	4176	&	0.52	&	1979.7246 	&	48800	&	6.3748	&		\\
201635132	& 1	&	 15.14	&	3983	&	0.44	&	1993.9755 	&	31550	&	5.8843	&	Noted in Osborn, et al. (2016)	\\
202071902	& 0	&	 10.40	&	6194    &	-	&	1958.5707	&	91224	&	5.3952	&		\\
202126877	& 0	&	 11.00	&	7142	&	-	&	1955.4341 	&	61834	&	5.8848	&		\\
202072917	& 0	&	 11.90	&	9221	&	-	&	1953.6261	&	248395	&	7.3536	&	Deep	\\
202060921	& 0	&	 12.00	&	7226	&	-	&	1971.3294	&	105768	&	17.8456	&	Deep	\\
202137580	& 0	&	 13.10	&	3755	&	-	&	1962.6458 	&	66511	&	50.9976	&	Deep; secondary eclipse	\\
202067195	& 0	&	 14.30	&	6980	&	-	&	1955.7918	&	15372	&	8.3376  &		\\
202135247	& 0	&	 14.40	&	3919	&	-	&	1963.8721 	&	89100	&	40.2096	&	Flat-bottom	\\
202073476	& 0	&	 15.00	&	4066	&   -	&	1957.7729 	&	47716	&	56.3904	&		\\
202085278	& 0	&	 15.50	&	3676	&	-	&	1948.2010 	&	8248	&	19.6128 &	\\

\enddata

\end{deluxetable}

\clearpage

 \end{landscape}



\end{document}